\documentclass[12pt,fleqn]{article}
\usepackage{amsmath,amssymb,stmaryrd,mathrsfs,latexsym,textcomp,graphicx}
\usepackage[usenames]{color}

\usepackage[top=2.2cm,bottom=2.2cm,left=2.2cm,right=2.2cm]{geometry}
\DeclareMathAlphabet{\mathpzc}{OT1}{pzc}{m}{it}

\begin{document}  
\title{\textbf{Neglected heavy leptons at the LHC}}
\author{B.~Holdom and M.~Ratzlaff\\
\emph{\small Department of Physics, University of Toronto}\\[-1ex]
\emph{\small Toronto ON Canada M5S1A7}}
\date{}
\maketitle
\begin{abstract}
New heavy leptons with standard model gauge couplings have well-determined cross sections for pair production. A standard pattern of mass mixing implies that the most likely decays are $\tau^\prime\to W \nu'$ and $ \nu'\to W \tau$.  Interestingly there have been no direct searches for heavy leptons with these decays at the LHC. However comparison with several multilepton searches allows us to set new limits on the heavy lepton masses. Three observed excesses in the signal regions prevent us from setting stronger limits.
\end{abstract} 
\subsection*{Introduction}
The discovery of a Higgs particle at the LHC \cite{Aad:2012tfa, Chatrchyan:2012ufa} puts strong indirect constraints on the existence of new heavy fermions. A new charged lepton adds to the $H\to\gamma\gamma$ amplitude a contribution that is $3\over4$ times the top contribution, which thus increases the destructive interference with the $W$ contribution. The $H\to\gamma\gamma$ width is then $\approx 0.64$ times smaller, but this result assumes that the Higgs particle has standard couplings to both the new lepton and the top and that there are not other degrees of freedom contributing to the loop as well. To avoid dependence on such assumptions it is preferable to have direct limits rather than apparent indirect limits. Also, a theory with a new sequential family of leptons and quarks where these assumptions are not correct and where a Higgs-like scalar can have Higgs-like widths has been presented in \cite{Holdom:2014bla}.

We choose to focus on the case of an extra chiral doublet of leptons $\left(\tau^\prime,\nu^\prime\right)$ with standard model gauge couplings. Since the $\nu'$ has a mass that is many orders of magnitude larger than the light neutrinos, its overlap with the flavor eigenstates of the light neutrinos should be extremely small. This is unlike the mass mixing between the three light neutrinos that can be responsible for the large mixing angles observed in the leptonic charged currents. The charged current mixing between $\tau^\prime$ or $\nu^\prime$ and the lighter leptons on the other hand must be due to the mass mixing in the charged lepton sector. Here we expect the $\tau^\prime$ flavor eigenstate to have its largest mass mixing with the $\tau$ flavor eigenstate, since this would mimic the mixing pattern in the similarly hierarchical quark sector. Then at least one of $\left(\tau^\prime,\nu^\prime\right)$ has a dominant decay to a third family lepton. For example if $m_{\tau^\prime}>m_{\nu^\prime}$ and if there is a sufficiently large mass splitting and/or a sufficiently small mixing angle then the dominant decays are $\tau^\prime\rightarrow W \nu^\prime$ and $ \nu^\prime\rightarrow \tau W$.  Pair production of heavy leptons then produces many different multilepton ($e$ or $\mu$) final states with kinematics quite different from current heavy lepton searches.
 
The current mass limits from LEP are 80.5 GeV (90.3 GeV) for a Majorana (Dirac) $\nu'$ decaying to $\tau W$ and 101.9 GeV (or 100.8 GeV) for $\tau'$ decaying to $\nu' W $ (or $\nu W$) \cite{Achard:2001qw}. We will be able to use LHC data to exclude a range of mass combinations above these limits. This is in lieu of a dedicated search for the simplest leptonic extension of the standard model, which surprisingly still remains to be done. An even simpler though less motivated search is possible if one assumes that $\nu'\rightarrow  W\ell$ ($\ell=\mu,e$) is the dominant decay. This was considered in \cite{Carpenter:2010bs} where it was found that a $\tau'$ mass up to at least 250 GeV would be excluded very early with 1 fb$^{-1}$ of data. We estimate that the corresponding limit with present data would be at least 600 GeV. Here again no dedicated search has been reported. Exotic vectorlike leptons have both nonstandard production and decay modes and constraints on such leptons as inferred from LHC multilepton searches are presented in \cite{Falkowski:2013jya}. Vectorlike leptons may also decay predominately to $\tau$s \cite{delAguila:2010es}, but the decays typically also involve $Z$ and $H$.
 
\subsection*{Production and decay channels} 
We first consider the case $m_{\tau^\prime}>m_{\nu^\prime}$ and with the decays already mentioned we have the following processes:
\begin{equation}\begin{split}
\mbox{production through $W$ }&\quad pp\rightarrow\tau^\prime \nu'\rightarrow\tau\tau WWW,\\
\mbox{neutrino pair production }&\quad pp\rightarrow \nu' \nu'\rightarrow\tau\tau WW,\\
\mbox{lepton pair production }&\quad pp\rightarrow {\tau^\prime} {\tau^\prime}\rightarrow\tau\tau WWWW.\\
\end{split}
\label{production}
\end{equation}
We do not consider the single production of a heavy lepton since the production cross section depends on an unknown mixing angle. At $\sqrt{s}=8$ TeV with $m_{\nu'}=100$ GeV  and $m_{ \tau^\prime}=150$ GeV,  the three pair production cross sections are $\sigma_{ \tau^\prime \tau^\prime}\sim170$ fb,  $\sigma_{ \nu'\nu'} \sim 260$ fb and $\sigma_{ \tau^\prime\nu'}\sim 690$ fb. We include an additional $K$ factor of $1.2$, as indicated by including one-loop corrections in Sherpa \cite{Gleisberg:2008ta}.  The cross sections are in the 150- to 4600-fb range for the range of heavy lepton and neutrino masses we consider in this analysis.

The $\nu'$ mass can be either Majorana or Dirac or some mixture. The pure Majorana case arises when the $\nu^\prime_R$ state is absent.\footnote{A Majorana mass produces a negative contribution \cite{Holdom:1996bn} to the electroweak parameter $T$, which may be helpful to offset other new physics (typically positive) contributions. A 150 GeV mass produces a shift in  $T$ of at least $\approx-0.5$.} In this case the $\nu'$ can decay via $ \nu'\rightarrow W^+\tau^-$ or  $ \nu'\rightarrow W^-\tau^+$ and so the decay chains are
 \begin{equation}
\begin{split}
p p& \rightarrow W^{\pm}\rightarrow {\tau^{\prime}}^\pm \nu'\rightarrow W^{\pm} \nu' \nu'\\
&\rightarrow\frac{1}{2}W^{\pm}W^-\tau^+W^+\tau^-+\frac{1}{4}W^{\pm}W^+\tau^-W^+\tau^-+\frac{1}{4}W^{\pm}W^-\tau^+W^-\tau^+\\
p p& \rightarrow Z,\gamma\rightarrow {\tau^{\prime}}^+ {\tau^{\prime}}^-\rightarrow W^{+}W^{-}\nu^\prime\nu^\prime\\
&\rightarrow\frac{1}{2} W^{+}W^{-}W^-\tau^+W^+\tau^-+\frac{1}{4} W^{+}W^{-}W^+\tau^-W^+\tau^-+\frac{1}{4} W^{+}W^{-}W^-\tau^+W^-\tau^+,\\
p p &\rightarrow Z\rightarrow {\nu^{\prime}}\nu^\prime\\
&\rightarrow\frac{1}{2}W^-\tau^+W^+\tau^-+\frac{1}{4}W^+\tau^-W^+\tau^-+\frac{1}{4}W^-\tau^+W^-\tau^+.
\end{split}
\label{decays}
\end{equation}
We find the branching fractions to various leptonic final states as shown in Table \ref{bf LN}. The most promising channels for detection have three or four leptons since these channels have small backgrounds.
 \begin{table}[hbtp]
\begin{center}
\begin{tabular}{|c|c|c|c|}
\hline
Decays&Branching fraction&Expected signal channel\\
\hline
$p p \rightarrow {\tau^{\prime}}^\pm \nu' \rightarrow \ell$+X&$0.39$ &$\ell+E^{\rm miss}_T+6$ or $7$ jets\\
$p p \rightarrow {\tau^{\prime}}^\pm \nu' \rightarrow \ell\ell$+X&$0.29$ &$\ell\ell+E^{\rm miss}_T+4,5$ or $6$ jets\\
$p p \rightarrow {\tau^{\prime}}^\pm \nu'\rightarrow \ell\ell\ell$+X&$0.10$ &$\ell\ell\ell+E^{\rm miss}_T+2,3$ or $4$ jets\\
$p p \rightarrow {\tau^{\prime}}^\pm \nu' \rightarrow \ell\ell\ell\ell$+X&$0.018$ &$\ell\ell\ell\ell+E^{\rm miss}_T+1$ or $2$ jets\\
\hline
$p p \rightarrow {\tau^{\prime}}^+ {\tau^{\prime}}^- \rightarrow \ell$+X&$0.35$ &$\ell+E^{\rm miss}_T+8$ or $9$ jets\\
$p p \rightarrow {\tau^{\prime}}^+ {\tau^{\prime}}^- \rightarrow \ell\ell$+X&$0.31$ &$\ell\ell+E^{\rm miss}_T+6,7$ or $8$ jets\\
$p p \rightarrow {\tau^{\prime}}^+ {\tau^{\prime}}^- \rightarrow \ell\ell\ell$+X&$0.17$ &$\ell\ell\ell+E^{\rm miss}_T+4,5$ or $6$ jets\\
$p p \rightarrow  {\tau^{\prime}}^+ {\tau^{\prime}}^- \rightarrow \ell\ell\ell\ell$+X&$0.037$ &$\ell\ell\ell\ell+E^{\rm miss}_T+2,3$ or $4$ jets\\
\hline
$p p \rightarrow  {\nu^{\prime}}\nu^\prime \rightarrow \ell$+X&$0.43$ &$\ell+E^{\rm miss}_T+4$ or $5$ jets\\
$p p \rightarrow  {\nu^{\prime}}\nu^\prime \rightarrow \ell\ell$+X&$0.25$ &$\ell\ell+E^{\rm miss}_T+2,3$ or $4$ jets\\
$p p \rightarrow  {\nu^{\prime}}\nu^\prime \rightarrow \ell\ell\ell$+X&$0.064$ &$\ell\ell\ell+E^{\rm miss}_T+1$ or $2$ jets\\
$p p \rightarrow  {\nu^{\prime}}\nu^\prime \rightarrow \ell\ell\ell\ell$+X&$0.006$ &$\ell\ell\ell\ell+E^{\rm miss}_T$\\
\hline
\end{tabular}
\end{center}
\caption{Branching fractions and decay products ($\ell=\mu,e$).}
\label{bf LN}
\end{table}

In the case where the $\nu^\prime$ mass is Dirac, its decay no longer produces $\tau$s of both signs and its pair production cross section is roughly twice the size of the Majorana $\nu^\prime$. However the $\nu^\prime\nu^\prime$ branching fractions to four leptons are small and so the increase in total number of four-lepton events would hardly be significant. The increase in three-lepton events is larger, but on the other hand for $\tau^\prime\nu^\prime$ production there is an increase in the branching fraction to an opposite sign same flavor (ossf) lepton pair and the three-lepton searches we consider have cuts related to ossf pairs. In any case the somewhat larger signals for the Dirac case means that the Majorana case provides limits for both cases.

If $m_{\nu^\prime}>m_{\tau^\prime}$ then the decays are $\nu^\prime\rightarrow\tau^\prime W$ and $\tau^\prime \rightarrow\nu_\tau W$. This has the effect that $\tau$s are replaced by $\nu_\tau$s in the final states. The four-lepton signal is much smaller since it only comes from neutrino pair production in this case, but the three-lepton signal can still be substantial.

Thus far we have assumed that the mixing angle that allows direct decays to $\tau$ or $\nu_\tau$ is very small. For a larger mixing angle, for instance $\sim\sqrt{m_\tau/m_{\tau^\prime}}\sim0.1$, the branching fractions of the heavier lepton are a sensitive function of $| m_{\tau^\prime}-m_{\nu^\prime}|$ when this is smaller than $m_W$. For small enough $| m_{\tau^\prime}-m_{\nu^\prime}|$ the decay $\tau^\prime\to W\nu_\tau$ can dominate when $m_{\tau^\prime}>m_{\nu^\prime}$, and the decay  $\nu^\prime\to W\tau$ can dominate when $m_{\tau^\prime}<m_{\nu^\prime}$. In the first case the branching to four leptons is the result of $\nu^\prime\nu^\prime$ decays only  and the signal is greatly diminished. The three-lepton branching fraction is also reduced. In the second case the branching fraction to three leptons is somewhat larger since $\tau$s have a larger leptonic branching fraction than $W$s. In the following we give results both for when the mixing angle is very small and when it is 0.1. The former corresponds to the limit when the heavier lepton decays to the lighter of the new leptons for any nonvanishing mass difference.
 
\subsection*{LHC multilepton searches}
 
Usually multilepton signals are considered in searches for supersymmetric or other exotic particles which then influences the event selection criteria. When colored particles decay to leptons the requirements related to the expected high jet $p_T$ reduce sensitivity to our signal since the jets in our final states are somewhat soft. High missing energy cuts and lepton $p_T$ requirements also reduce the sensitivity, especially for smaller neutrino masses and/or mass difference. An example of a low sensitivity search would be the CMS search for same-sign leptons and jets \cite{Chatrchyan:2013fea} where the combination of cuts on the scalar sum $H_T^j$ of transverse jet momentum and on the missing energy $E_T^{\rm miss}$ reduce our signal to approximately the size of the error on the background. Another example is the $H\to 4\ell$ \cite{Aad:2012tfa,Chatrchyan:2012ufa} search where the lepton $p_T$ and invariant mass selections reduce our signal to a very small number of events in the search region.
 
There are dedicated searches for exotic heavy leptons such as seesaw triplet leptons $(\Sigma^\pm,\Sigma_0)$ produced via $pp\to \Sigma^\pm\Sigma_0$. Here ATLAS and CMS usually assume that the triplet lepton masses are degenerate and that the heavy leptons decay predominately to light leptons \cite{CMS:2012ra,ATLAS:2013hma} (as may also be applicable to composite models \cite{Redi:2013pga}). In our case these searches may only be marginally sensitive to our final states. CMS also has a search for the single production of a heavy Majorana neutrino via $pp\to W\to N\ell$  ($\ell=\mu,e$) \cite{Chatrchyan:2012fla}, but this production cross section is sensitive to an unknown and likely very small mixing angle. The heavy right-handed neutrino search \cite{CMS:2012uaa} has high lepton $p_T$ cuts and invariant mass cuts that are not appropriate for our signal. None of these searches have explored the case where the dominate decay of the heavy neutrino is $\nu' \to \tau W$. 
 
There are searches with moderate cuts and small enough backgrounds that are sensitive to our signal; in order of increasing sensitivity there are the searches for Higgs associated production where $HW\to 3\ell 3\nu$ \cite{CMS:zwa,TheATLAScollaboration:2013hia}, the ATLAS anomalous three-lepton search \cite{TheATLAScollaboration:2013cia}, the CMS anomalous three or more leptons search \cite{Chatrchyan:2014aea}, and the ATLAS four or more leptons SUSY search \cite{ATLAS:2013qla}. For each search we outline which cuts have been considered and then estimate our signal based on those cuts for a range of heavy lepton mass combinations. By comparing to the data we can determine the excluded mass combinations.
   
We use Herwig$++$ 2.7.0 \cite{Bahr:2008pv} interfaced with a Feynrules \cite{Alloul:2013bka} new physics model to handle the production, decay and showering of the events.  We feed the fully showered events through the detector simulator Delphes $3$ \cite{deFavereau:2013fsa}. We use the default ATLAS and CMS Delphes cards modified with the lepton isolation cuts appropriate to the particular searches we consider. Then Madanalysis $5$ \cite{Conte:2012fm} is used to incorporate the key kinematic cuts to allow us to compare with the experimental results.
   
The program Checkmate \cite{Drees:2013wra} provides a streamlined framework to compare model predictions against a certain set of validated ATLAS and CMS analyses, although none of the searches we investigate are in this set. Included in this framework is a version of Delphes 3 modified with various efficiency formulas that more closely match the characteristics of the two experiments. We were able to slightly modify the Checkmate Delphes to interface it with Madanalysis 5 and to use the lepton isolation cuts appropriate to our searches. The two versions of Delphes gave quite similar results, except as noted below. In the cases they differ we adopt the implied normalization from Checkmate.

\subsubsection*{Higgs associated production $HW\to \ell\ell\ell+E_T^{\rm miss}$ \cite{CMS:zwa,TheATLAScollaboration:2013hia}}
The CMS search \cite{CMS:zwa} puts limits on the $HW$ cross section with the 8 TeV 19.5 fb$^{-1}$ data set. The $3\ell$ channel with no ossf pairs has a small background and is the most sensitive to our signal. For this channel we consider the following selection criteria:
\begin{itemize}
\item The leading lepton has $p_T^{\ell_1}>20$ GeV and additional leptons have $p_T^{\ell}>10$ GeV.
\item The sum of charge of the three leptons is $\pm 1$.
\item The smallest invariant mass of two leptons is $m_{\ell\ell}  <100$ GeV and the smallest distance between oppositely charged leptons is $\Delta R_{\ell^+\ell^-}<2$.
\item Missing energy $E_T^{\rm miss}>30$ GeV.
\item A jet veto (no jet with $E_T^j>40$ GeV).
\end{itemize}
As a check on our results we use Herwig++ to estimate the number of events for $HW$ production satisfying the selection criteria above. In the $HW\to WWW$ channel our estimate  is $0.68\pm0.14$ as compared to $0.92\pm0.16$ from CMS, so our limits can be considered conservative.
\begin{figure}[h]	
\centering
\includegraphics[scale=0.35]{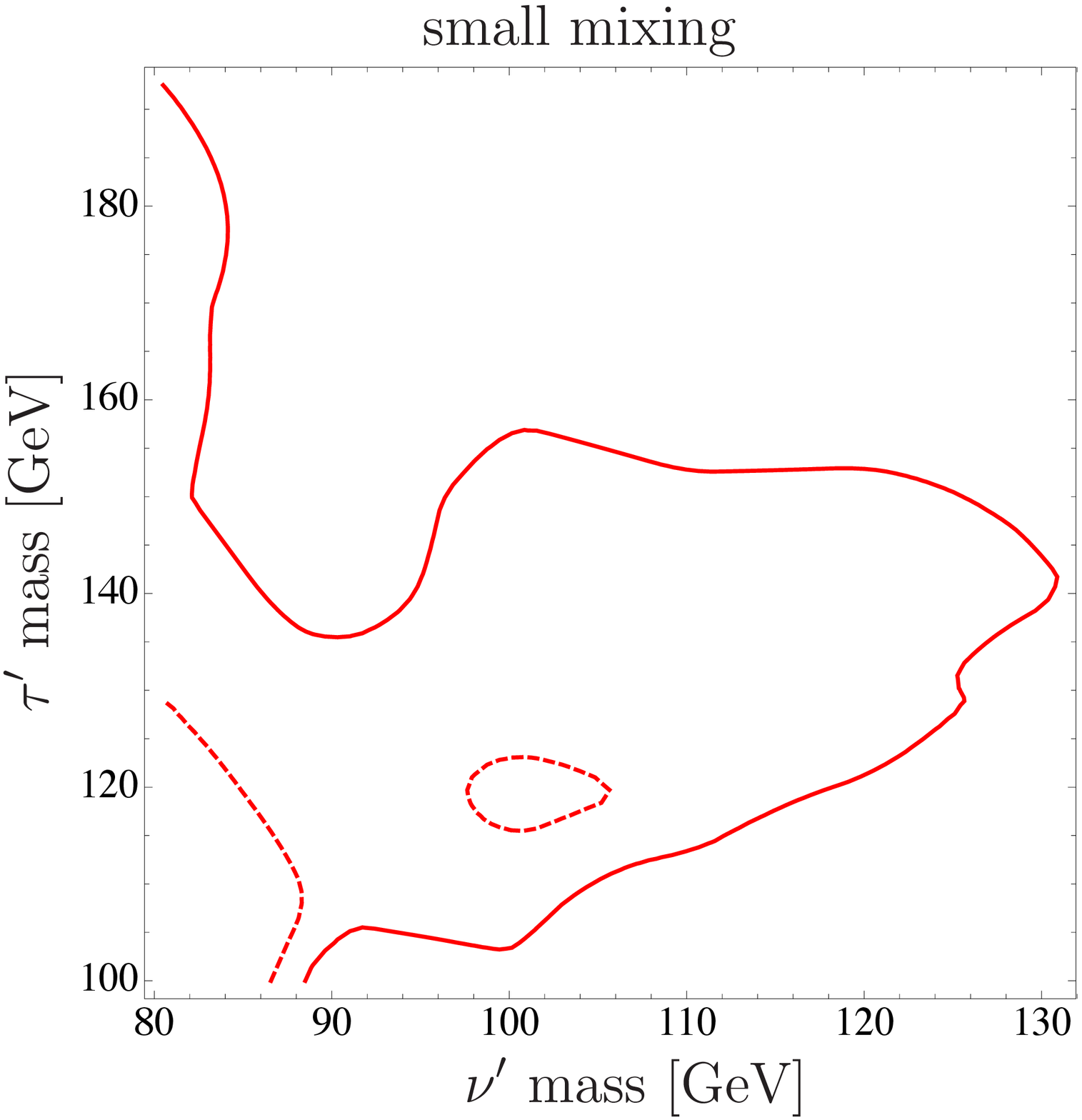}\quad\quad
\includegraphics[scale=0.35]{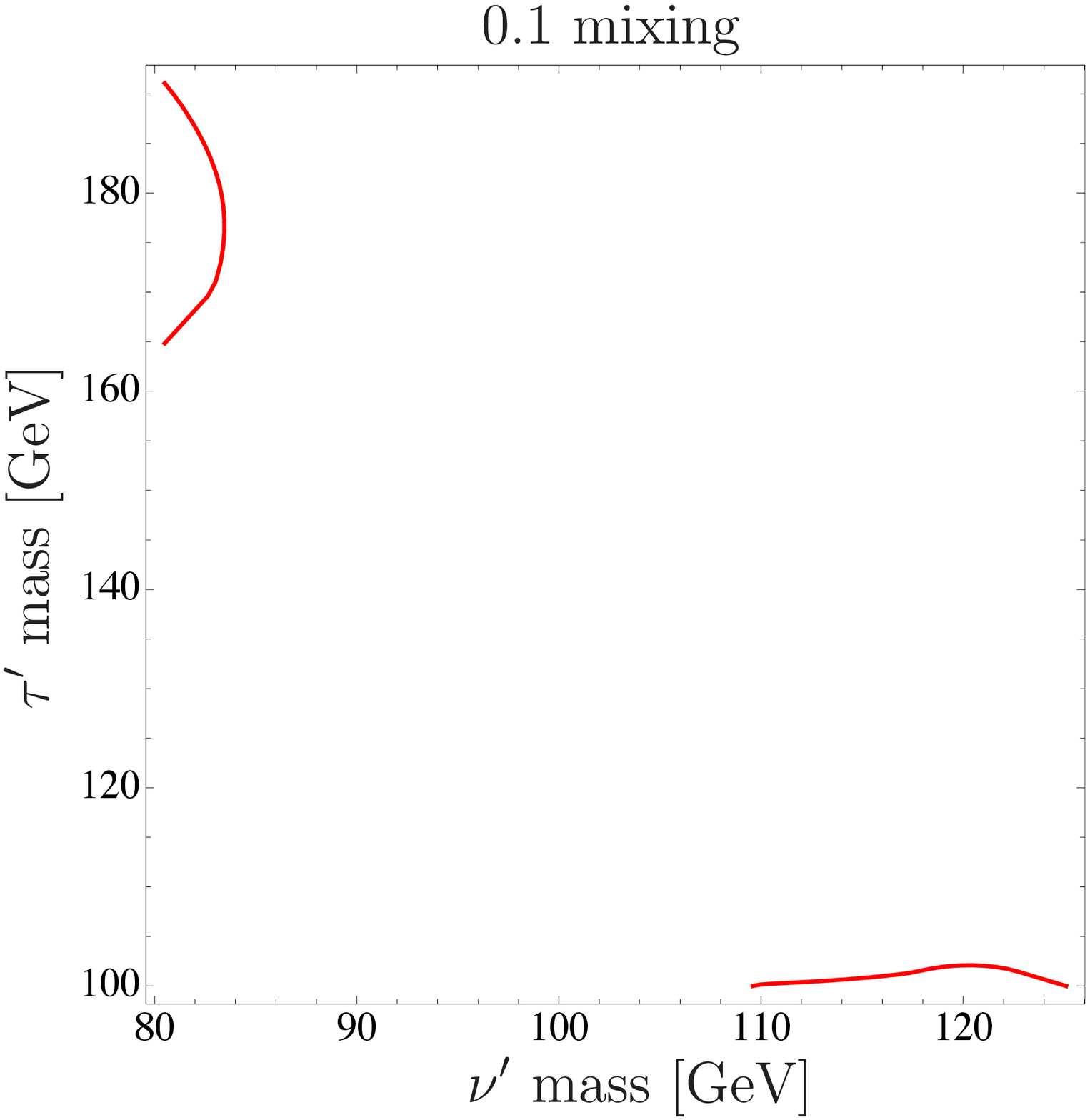}
\caption{The $\tau^\prime$ and $\nu'$ mass combinations within the solid (dashed) line are excluded at $95\%$ C.L.~($99\%$ C.L.) from the CMS $HW$-associated production search \cite{CMS:zwa}. }
\label{exclusion HW}
\end{figure}

The number of observed events in this channel is consistent with the expected number of events including SM-sized $HW$ production. An approximate limit on the number of excess events that are allowed is 4.2 events (6.9 events) at 95\% C.L.~(99\% C.L.). The resulting ranges of mass combinations that are excluded are shown in Fig.~\ref{exclusion HW}, for the cases of small and large (0.1) mixing angle. For the small mixing case the majority of excluded points is in the region where the mass difference between the $\tau' $ and the $\nu'$ is less than $m_W$.  For 0.1 mixing the branching fraction to three leptons is reduced in this region and we would expect only approximately two or three events.

We mention that for this analysis a serious discrepancy occurred between Checkmate (the results of which we have quoted) and the original Delphes. The latter gives a signal estimate approximately three times larger\footnote{Events that pass the jet veto have a lower average missing energy when generated with Checkmate than with Delphes. There is also a similar effect for events that pass the lepton selection. Thus the combination of jet veto and lepton selection results in significantly fewer events passing the missing energy cut when generated with Checkmate.} and so here again our limits can be seen as conservative.

The similar ATLAS search \cite{TheATLAScollaboration:2013hia} has somewhat more relaxed jet requirements and so potentially this search could be more sensitive to our signal. But the equivalent channel of interest, the $Z$-depleted channel, has an excess (nine events observed, $3.6\pm0.5$ events expected). We discuss this further in the Conclusion.

\subsubsection*{ATLAS search for new physics in events with three charged leptons \cite{ TheATLAScollaboration:2013cia}}
This ATLAS search at $\sqrt{s}=8$ TeV and $20.3$ fb$^{-1}$ explores channels with either three light leptons ($3\ell$) or a $\tau$ and two light leptons. Only the $3\ell$ channel is sensitive to our signal, and the most inclusive of the various selection criteria considered by ATLAS proved to be the most sensitive. Thus for the $3\ell$ channel we consider the following selection criteria.
\begin{itemize}
\item  The leading lepton is required to have  $p_T^{\ell_1}>26$ GeV and additional leptons have $p_T^{\ell}>15$ GeV.
\item Lepton invariant mass $m_{\ell^+\ell^-}>15$ GeV.
\item A $Z$ veto removing events with ossf pairs having $m_{\ell^+\ell^-}$ within $m_Z\pm20$ GeV.
\end{itemize}
\begin{figure}[h]	
\centering
\includegraphics[scale=0.35]{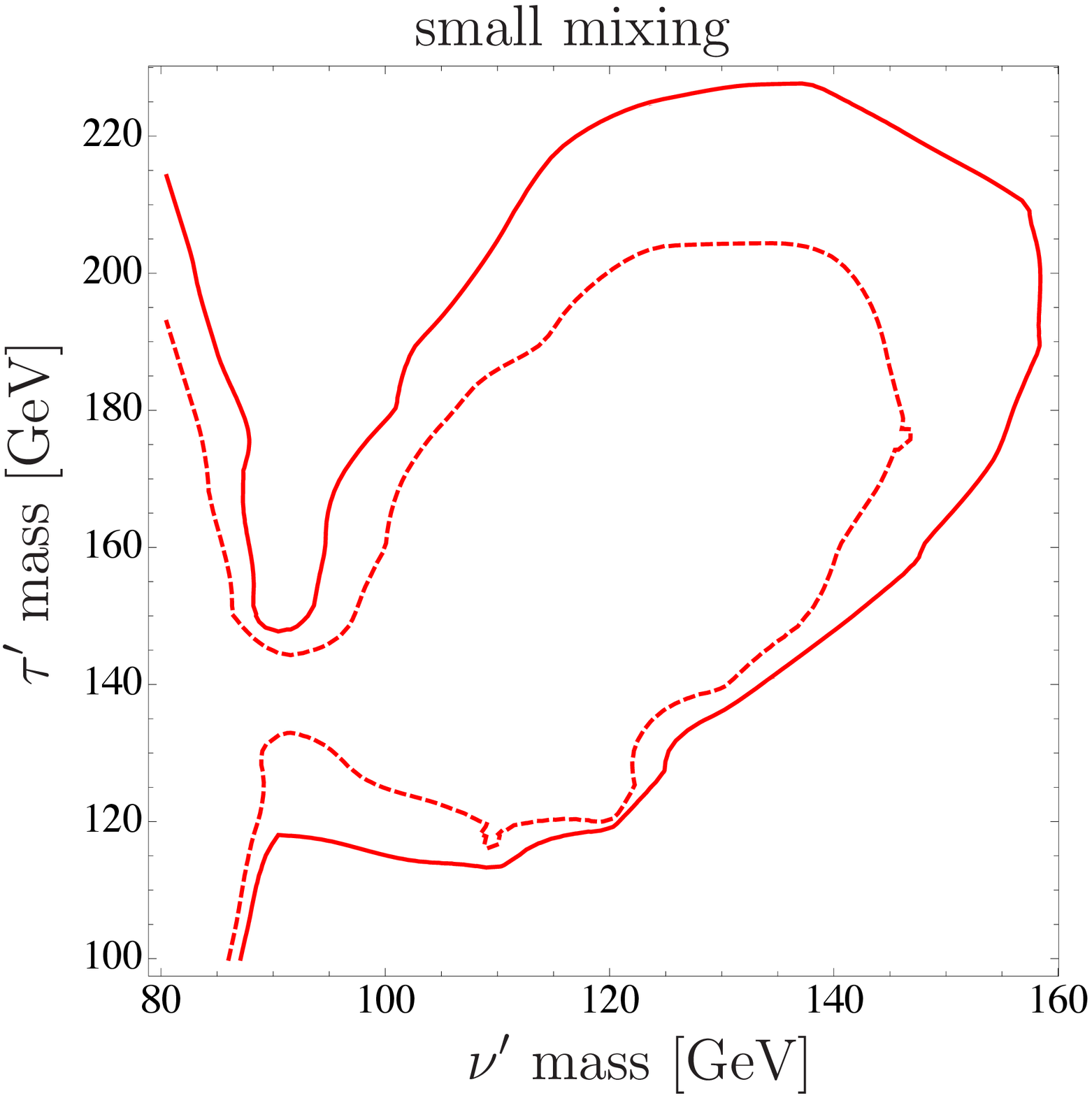}\quad\quad
\includegraphics[scale=0.35]{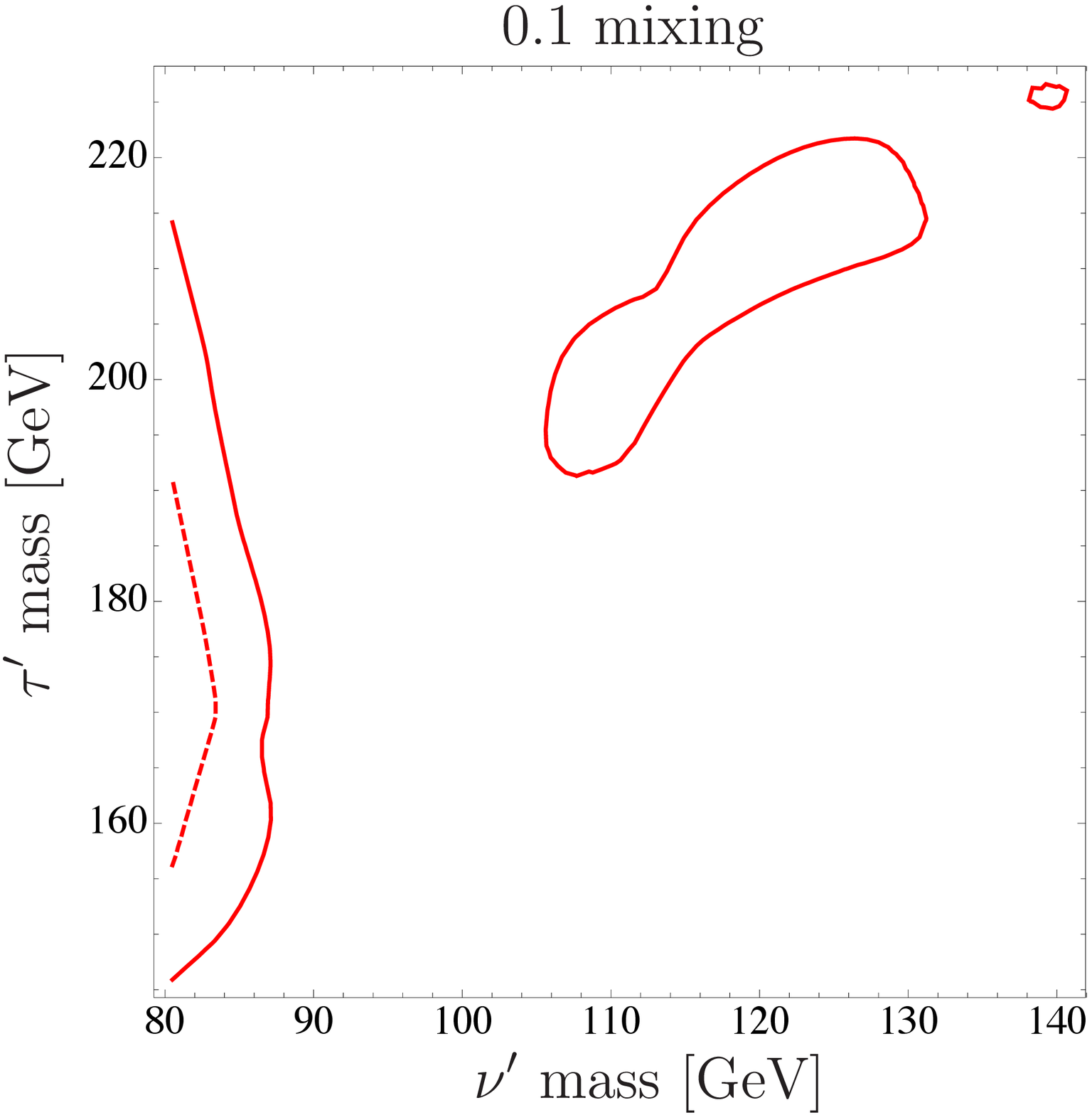}
\caption{The $\tau^\prime$ and $\nu'$ mass combinations within the solid (dashed) line are excluded at $95\%$ C.L.~($99\%$ C.L.) from the ATLAS anomalous three-lepton search \cite{TheATLAScollaboration:2013cia}.}
\label{3L}
\end{figure}

For small mixing angle (0.1 mixing angle) we obtain approximately 30--270 events (40--170 events) at the mass combinations we have explored. The 95\% C.L.~(99\% C.L.) exclusion is 110 events (135 events) and this yields the excluded mass combinations  shown in Fig.~\ref{3L}.  For 0.1 mixing most of the excluded region corresponds to $m_{\tau'}-m_{\nu'}>m_{W}$; however, in the $m_{\tau'}<m_{\nu'}$ region there are still 50--80 events, and so future similar searches may be sensitive to this region as well.

\subsubsection*{CMS search for anomalous production of events with three or more leptons \cite{Chatrchyan:2014aea}}
This multilepton search (at $8$ TeV with $19.5$ fb$^{-1}$)  focuses on  SUSY models where the final state may  also have jets and missing energy. The data is divided into many channels with three or four leptons and exclusive channels for combinations of $\ell=e,\mu$ and up to one hadronically decaying $\tau$ (also $b$s, but not of interest for our signal). The channels are further divided into $50$ GeV missing energy bins, scalar sum of the jet transverse momentum $H_T$ greater or less than 200 GeV, $Z$ inclusive or vetoed and by the number of ossf pairs. Our signal would not produce a significant number of events in any particular four-lepton channel, and for three leptons our signals would be on order the size of the background error.

Therefore some combination of the channels is needed to produce a favorable signal to background ratio. The combinations we choose have no missing energy requirements and the selection criteria they have in common are the following.
\begin{itemize}
\item For light leptons $\ell$, the leading lepton has  $p_T^\ell>20$ GeV and the remaining selected leptons have  $p_T^\ell>10$ GeV.
\item Tagged $\tau$s have $p_T^\tau>20$ GeV.
\item Lepton invariant mass $m_{\ell^+\ell^-}>12$ GeV.
\item A $Z$ veto removing events with ossf pairs having $m_{\ell^+\ell^-}$ within $m_Z\pm15$ GeV.
\end{itemize}
We choose to combine the $H_T$ low and $H_T$ high channels, although we do not expect a significant number of events in the high $H_T$ bins.
 \begin{figure}[h]	
\centering
\includegraphics[scale=0.35]{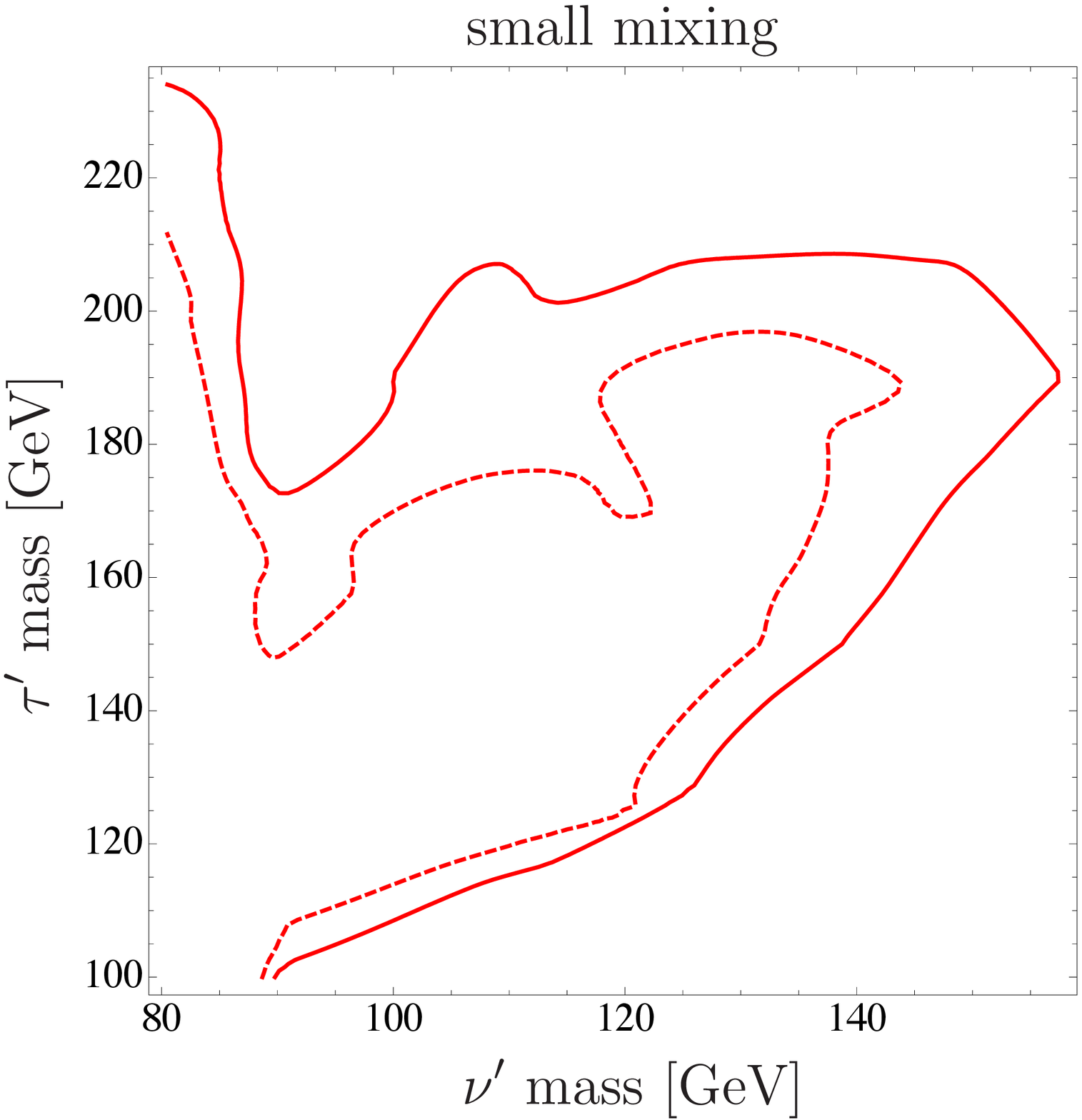}\quad\quad
\includegraphics[scale=0.35]{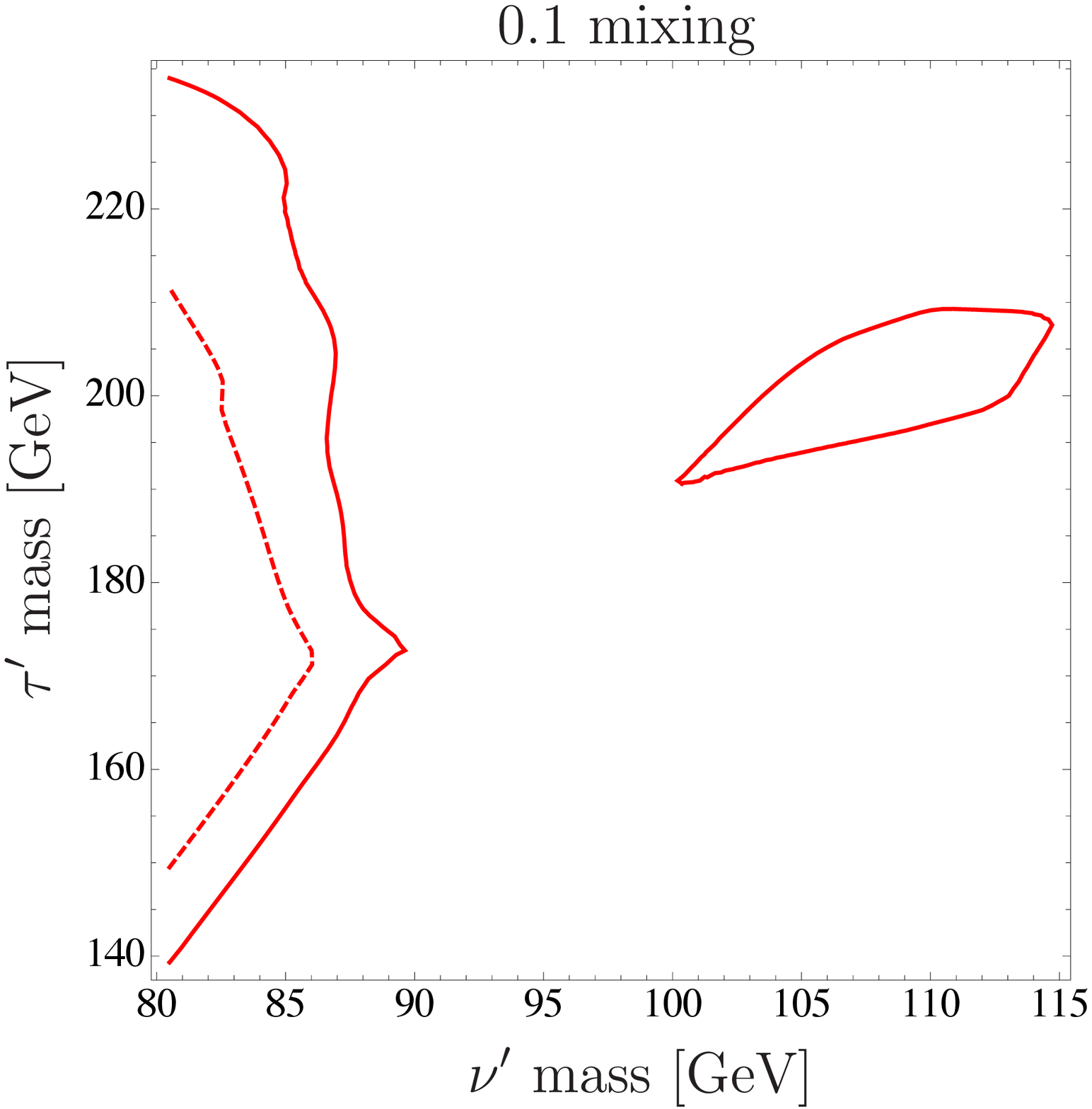}
\caption{The $\tau^\prime$ and $\nu'$ mass combinations within the solid (dashed) line are excluded at $95\%$ C.L.~($99\%$ C.L.) from the CMS anomalous three or more leptons search \cite{Chatrchyan:2014aea} in the three-lepton channels with no ossf pairs.}
\label{exclusion3LDY0}
\end{figure}

The first combination we consider has all the three-lepton channels with no ossf pairs and no tagged $\tau$s. Here our signal produces 16--150 events for small mixing and 16--80 events for 0.1 mixing. The expected number of background events in this channel is $109\pm 19$, and the observed number of events is $107$. From this the 95\% C.L.~(99\% C.L.) exclusion is 40 events (52 events). The excluded regions are shown in Fig.~\ref{exclusion3LDY0}.

A second combination has all the four light lepton channels excluding channels with two ossf pairs. This combination is better suited to our signal than the previous searches and so could be expected to put stronger limits on heavy lepton masses, but there is an excess; the expected background is $6\pm1$, while the observed number of events is $10$. 

A third combination has all of the four-lepton channels with one tagged $\tau$ and one ossf pair. For an estimate of our signal we must rely on Checkmate's model of the efficiencies for tagging $\tau$s. We find that this combination is also sensitive to our signal, but here again there is an excess; the expected background is $11\pm2$ events, while the observed number of events is $24$. This excess was also noted in \cite{D'Hondt:2013ula}. We comment more on these excesses in the Conclusion.

\subsubsection*{ATLAS search for supersymmetry in events with four or more leptons \cite{ATLAS:2013qla}} 
This multilepton search ($20.7$ fb$^{-1}$ at $\sqrt{s}=8$ TeV) has moderate cuts and small backgrounds so it is a good match for our signal. Of the two channels without $\tau$s, SR0noZa and SR0noZb, we focus on the first since it is somewhat more sensitive to our signal. We consider the following selection criteria.
\begin{itemize}
\item Four or more leptons ($e$ or $\mu$).
\item All leptons have $p_T^\ell>10$ GeV.
\item Lepton invariant mass $m_{\ell^+\ell^-}>12$ GeV.
\item  Missing energy $E_T^{\rm miss}> 50$ GeV.
\item An extended $Z$ veto for events with $m_{\ell\ell}$, $m_{3\ell}$ and $m_{4\ell}$ within $10$ GeV of the $Z$ mass.
\end{itemize}
\begin{figure}[h]	
\centering
\includegraphics[scale=0.35]{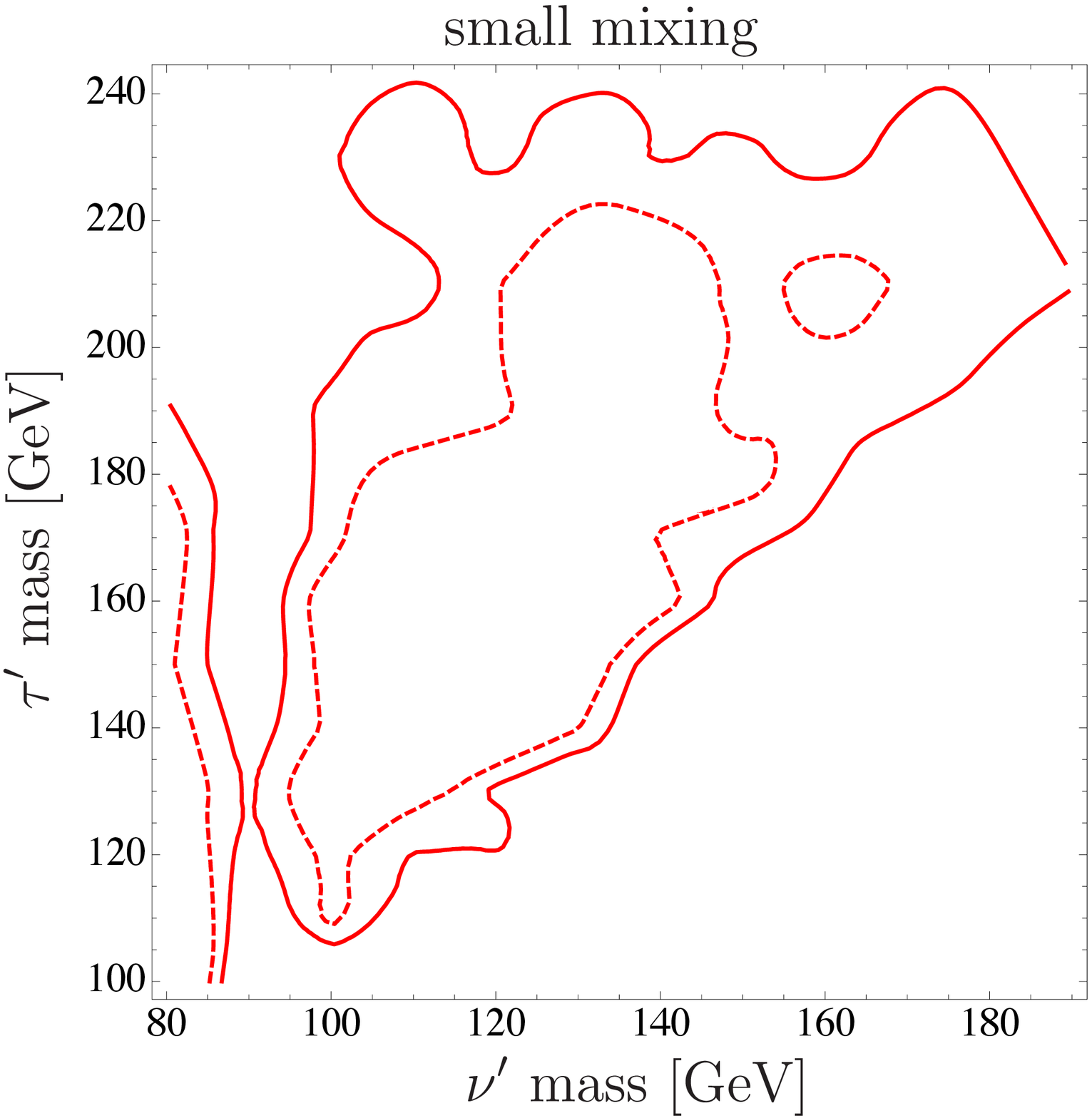}\quad\quad
\includegraphics[scale=0.35]{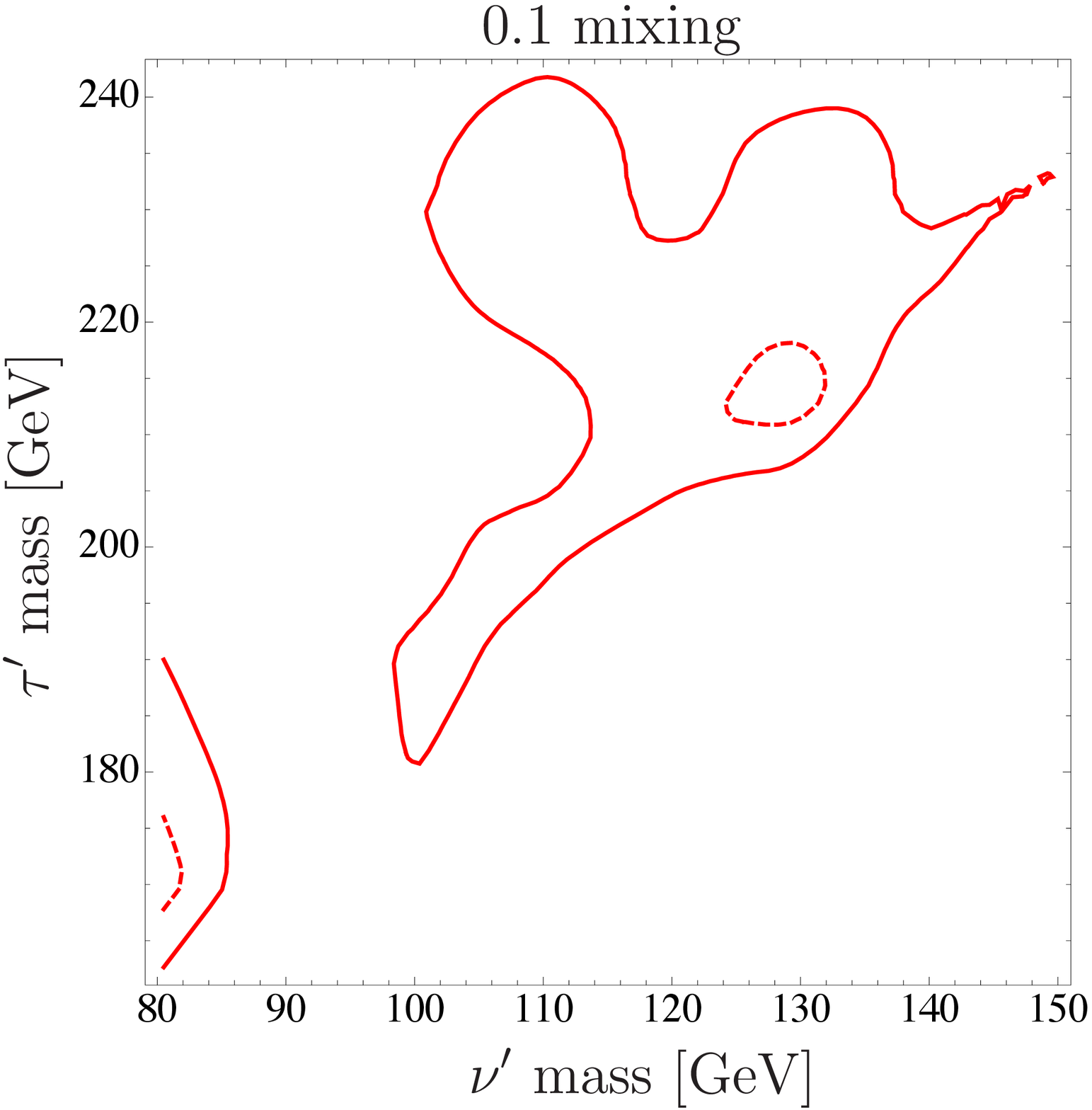}
\caption{The $\tau^\prime$ and $\nu'$ mass combinations within the solid (dashed) line are excluded at $95\%$ C.L.~($99\%$ C.L.) from the ATLAS four or more leptons SUSY search \cite{ATLAS:2013qla} in the SR0noZa channel.}
\label{exclusion1}
\end{figure}
\begin{figure}[h]	
\centering
\includegraphics[scale=0.35]{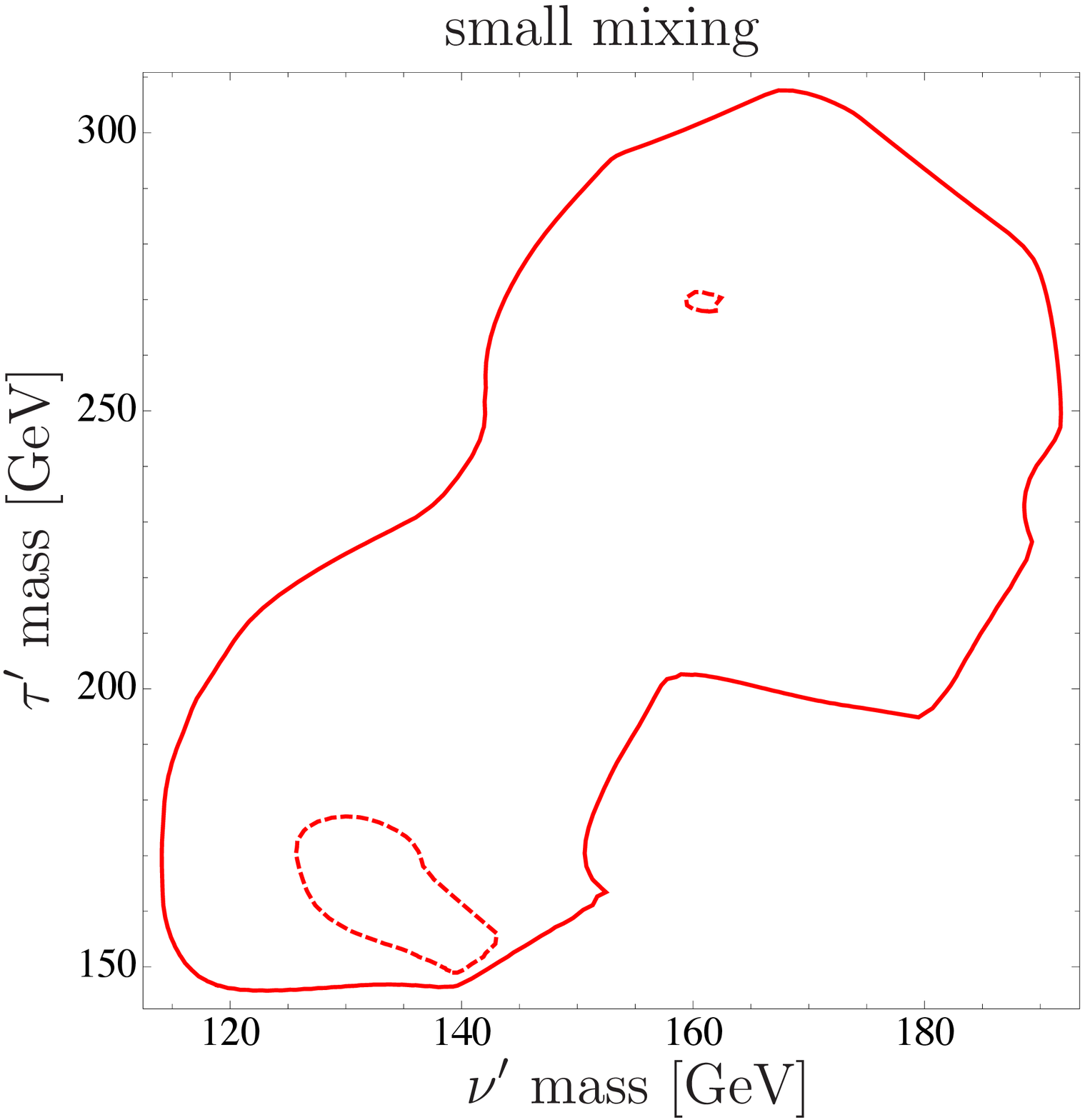}\quad\quad
\includegraphics[scale=0.35]{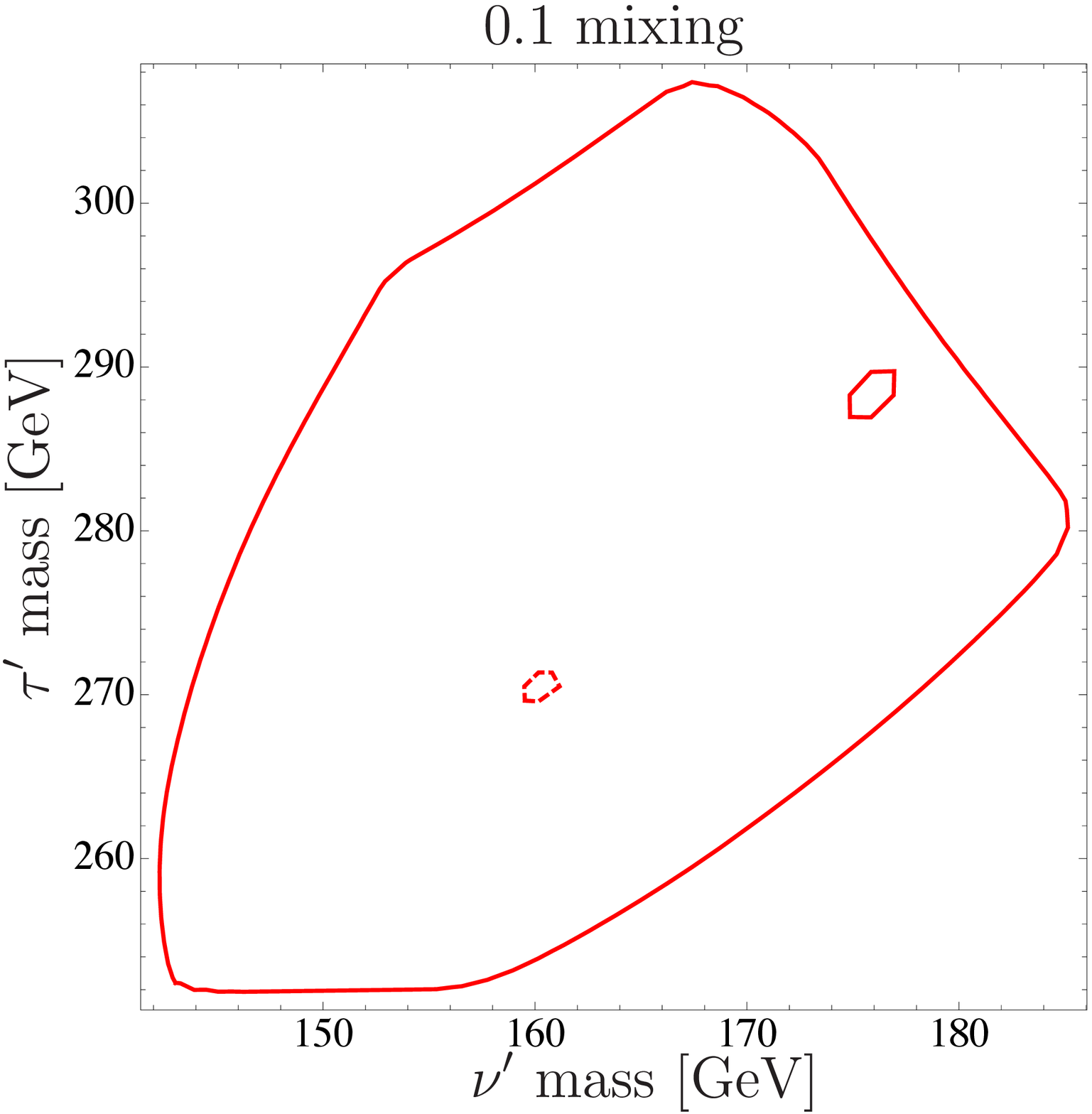}
\caption{The $\tau^\prime$ and $\nu'$ mass combinations within the solid (dashed) line are excluded at $95\%$ C.L.~($99\%$ C.L.) from the ATLAS four or more leptons SUSY search \cite{ATLAS:2013qla} in the SR1noZ channel.}
\label{exclusion2}
\end{figure}
Here Checkmate Delphes produced a 20\% increase in signal strength relative to the original Delphes. As a check of our signal strength estimate we generated events for the $WWZ$ background using Madgraph \cite{Alwall:2011uj} and PYTHIA \cite{Sjostrand:2006za}. We find that our estimate for this background, with the above selection criteria, is in agreement with the ATLAS estimate within error. 

The 95\% C.L.~(99\% C.L.) exclusion from the SR0noZa channel is 4.7 events (6.8 events) and this yields the excluded mass combinations  shown in Fig.~\ref{exclusion1}.  For $0.1$ mixing we see that the four-lepton branching fraction is only significant for $m_{\tau'} -m_{\nu'}\gtrsim m_W$.

Next we consider the search channel with a $\tau$, SR1noZ. The search criteria differing from the previous set are the following:
\begin{itemize}
\item Three leptons ($e$ or $\mu$).
\item A tagged $\tau$ with $p_T^\tau>20$ GeV.
\item  Missing energy $E_T^{\rm miss}> 100$ GeV or $m_{\rm eff}>400$ GeV.
\end{itemize}
$m_{\rm eff}$ is defined as the scalar sum of $E_T^{\rm miss}$, the $p_T$ of signal leptons and the $p_T$ of signal  jets with $p_T > 40$ GeV. To be conservative we adopt the ``tight'' identification criteria for $\tau$s in Checkmate. The 95\% C.L.~(99\% C.L.) exclusion from the SR1noZ channel is 7.5 events (9.9 events) and this yields the excluded mass combinations  shown in Fig.~\ref{exclusion2}.\footnote{The authors of Checkmate provided us with their unvalidated analysis code for this ATLAS search. This provides a signal estimate for SR0noZa that agrees well with ours, while the signal estimate for SR1noZ is only somewhat smaller than ours.}

Additionally we have checked two of the validation regions in this search for their sensitivity to our signal. Both of these regions have missing energy less than 50 GeV and an extended $Z$ veto. The first validation region (VR0noZ) has four or more leptons and our signal outside the mass region already excluded would be at most the size of the background error. The second validation region (VR1noZ) has three leptons and one $\tau$; here our signal would be at most eight events outside the excluded region. This does not enlarge the excluded region.\footnote{The VR1noZ validation region overlaps with the combination of the previous section with  one $\tau$ tag, and the former does not have an excess while the latter does. However the former region does not cover the missing energy range 50--100 GeV and $\sim1/2$ the CMS excess is in this region.}

\subsection*{Conclusion}
We have investigated the multilepton signals for a new heavy doublet of leptons with nonexotic quantum numbers. A standard pattern of mass mixing following that in the quark sector would mean that the new doublet mixes mostly with the third family leptons. The most accessible multilepton signatures occur when the dominant decays are $\tau^\prime\to \nu' W$ and $\nu'\to W \tau$. There is no dedicated search at the LHC for heavy leptons with these decays. Instead the heavy lepton searches have so far focused on more exotic decays such as charged lepton decays to $W \nu_\ell$ or $Z\ell$ ($\ell=e$ or $\mu$) for the most part, and heavy neutrino decays to $W\ell$. At this time these searches are not particularly sensitive to our signal.

\begin{figure}[h]	
\centering
\includegraphics[scale=0.44]{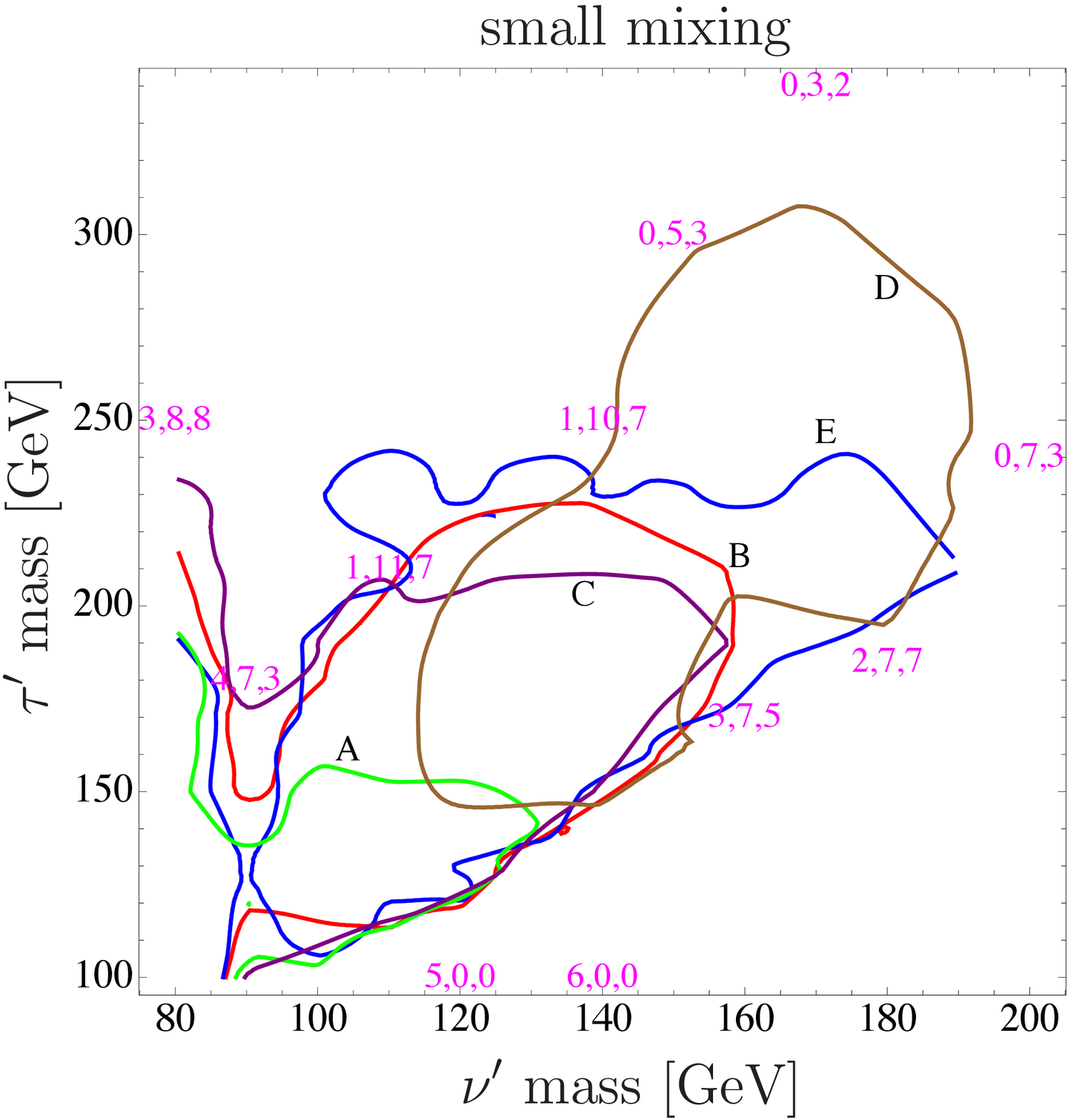}
\includegraphics[scale=0.44]{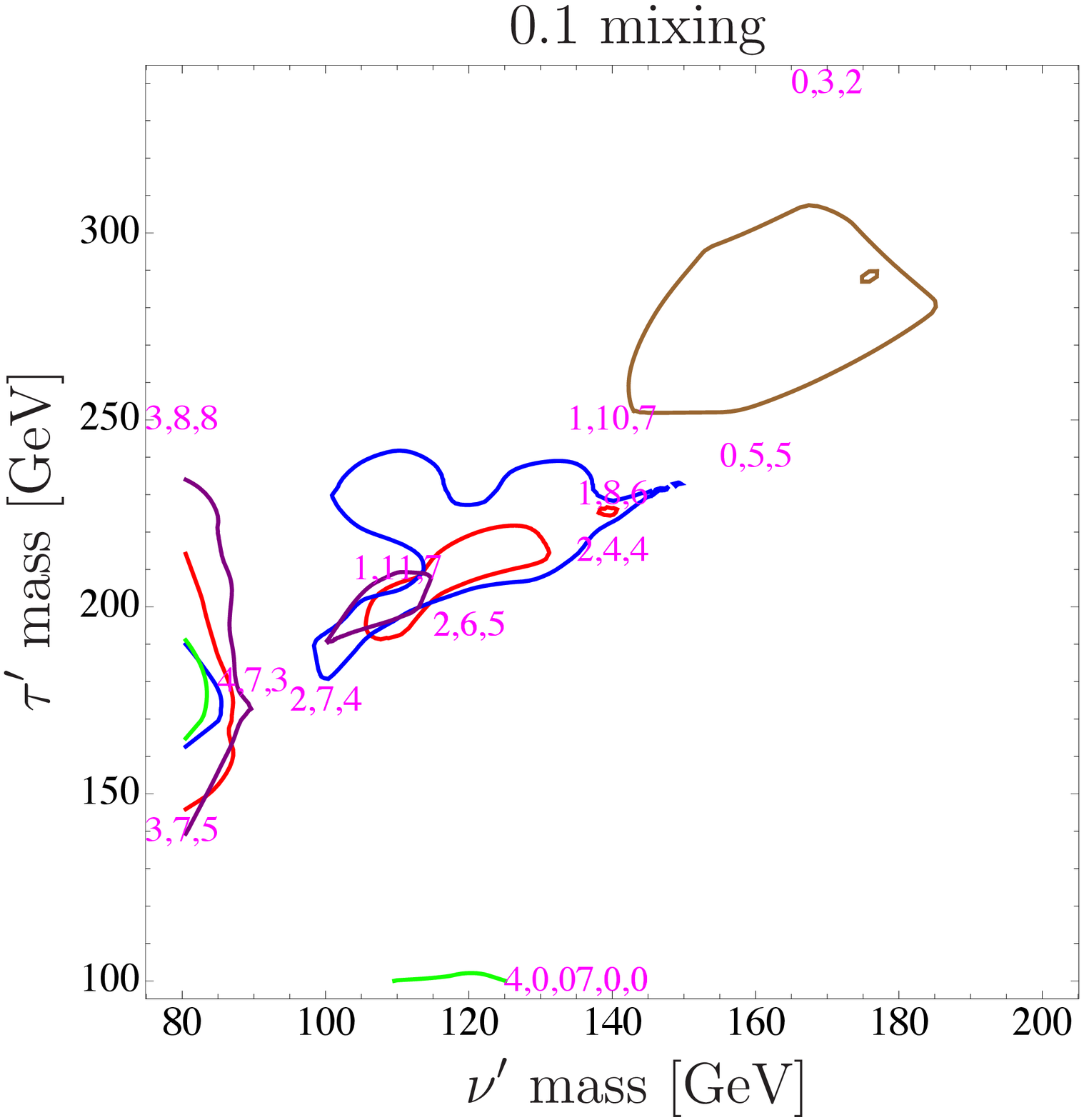}
\caption{A comparison of the exclusion regions. Line A (green) is excluded by the CMS $HW$ -associated production search \cite{CMS:zwa}. The B (red) is excluded by the ATLAS anomalous three-lepton search \cite{TheATLAScollaboration:2013cia}. Line C (purple) is excluded by the CMS anomalous three or more leptons search \cite{Chatrchyan:2014aea} in the three-lepton channel with no ossf pairs. Line D (brown) [line E (blue)] is excluded by the ATLAS four or more leptons SUSY search \cite{ATLAS:2013qla} with (without) at tagged $\tau$. At selected points we give the number of expected signal events for, from left to right, the ATLAS $HW$-associated production search \cite{TheATLAScollaboration:2013hia} and two four-lepton channels in the CMS anomalous three or more leptons search \cite{Chatrchyan:2014aea}, one without two ossf pairs and the other with one tagged $\tau$ and one ossf pair.  } 
\label{total}
\end{figure}

However there are multilepton searches that are not as model dependent and have moderate missing energy cuts and low lepton $p_T$ requirements that do put limits on our signal. These limits depend strongly on the mixing angle when the mass difference between the leptons is less than $m_W$. In Fig.~\ref{total} we collect together the excluded mass regions from the four searches we have considered that are able to set limits.

We have mentioned three excesses that have prevented stronger limits, one in the ATLAS $Z$-depleted three-lepton channel for $HW$ production and two among the four-lepton channels in the CMS anomalous three or more leptons search, without and with a $\tau$ tag, respectively. None of these excesses are individually very significant, but our signal would produce a small number of events in these searches while still remaining consistent with the limits we have obtained. This is indicated in Fig.~\ref{total} which shows the number of signal events at a selected set of mass combinations in each of these three searches.

A different origin for new multilepton events typically results in a different pattern of excesses. As an example we mentioned earlier heavy leptons that decay directly to $\ell=\mu\mbox{ or }e$. Since the lower limit on the masses of such leptons are constrained to be much higher, the jets are more energetic and less likely to pass the jet-related cut in the $HW$ search. In the four-lepton search the heavier leptons would populate the high $H_T>200$ GeV region where the background is very small, while most of the excess is in the low $H_T<200$ GeV region. Finally no excess would be expected in final states with a tagged $\tau$.

The searches we have considered are not completely optimized for our signal. In particular the ATLAS anomalous three-lepton search \cite{TheATLAScollaboration:2013cia} provides several additional selections after the inclusive selection, but the lowest cuts on the kinematic variables are too high for our signal. For example it would be more appropriate if the missing energy cut was 50 rather than 100 GeV. In the case of four leptons it may be that a search more optimized for our signal would have a loose $Z$ veto (such as no ossf pairs within $m_Z\pm10$ GeV) and a loose missing energy cut (such as $E_T^{\rm miss}>25$ GeV).
 
There have been quite extensive searches for new sequential quarks with decays to third family quarks and so the absence of searches for new sequential leptons with decays to third family leptons is odd. From our results it is clear that a dedicated multilepton search could close this gap, clear up the status of some excesses and further improve on the mass bounds for such leptons.

\section*{Acknowledgments}
This work was supported in part by the Natural Science and Engineering Research Council of Canada.

\end{document}